\documentstyle[twoside,psfig]{article}

\catcode`\@=11
\long\def\@makefntext#1{
\protect\noindent \hbox to 3.2pt {\hskip-.9pt  
$^{{\eightrm\@thefnmark}}$\hfil}#1\hfill}		

\def\thefootnote{\fnsymbol{footnote}}
\def\@makefnmark{\hbox to 0pt{$^{\@thefnmark}$\hss}}	
	
\def\ps@myheadings{\let\@mkboth\@gobbletwo
\def\@oddhead{\hbox{}
\rightmark\hfil\eightrm\thepage}   
\def\@oddfoot{}\def\@evenhead{\eightrm\thepage\hfil
\leftmark\hbox{}}\def\@evenfoot{}
\def\sectionmark##1{}\def\subsectionmark##1{}}

\oddsidemargin=\evensidemargin
\renewcommand{\thefootnote}{\fnsymbol{footnote}}

\newcounter{sectionc}\newcounter{subsectionc}\newcounter{subsubsectionc}
\renewcommand{\section}[1] {\vspace{12pt}\addtocounter{sectionc}{1} 
\setcounter{subsectionc}{0}\setcounter{subsubsectionc}{0}\noindent 
	{\tenbf\thesectionc. #1}\par\vspace{5pt}}
\renewcommand{\subsection}[1] {\vspace{12pt}\addtocounter{subsectionc}{1} 
	\setcounter{subsubsectionc}{0}\noindent 
	{\bf\thesectionc.\thesubsectionc. {\kern1pt \bfit #1}}\par\vspace{5pt}}
\renewcommand{\subsubsection}[1] {\vspace{12pt}\addtocounter{subsubsectionc}{1}
	\noindent{\tenrm\thesectionc.\thesubsectionc.\thesubsubsectionc.
	{\kern1pt \tenit #1}}\par\vspace{5pt}}
\newcommand{\nonumsection}[1] {\vspace{12pt}\noindent{\tenbf #1}
	\par\vspace{5pt}}

\newcounter{appendixc}
\newcounter{subappendixc}[appendixc]
\newcounter{subsubappendixc}[subappendixc]
\renewcommand{\thesubappendixc}{\Alph{appendixc}.\arabic{subappendixc}}
\renewcommand{\thesubsubappendixc}
	{\Alph{appendixc}.\arabic{subappendixc}.\arabic{subsubappendixc}}

\renewcommand{\appendix}[1] {\vspace{12pt}
        \refstepcounter{appendixc}
        \setcounter{figure}{0}
        \setcounter{table}{0}
        \setcounter{lemma}{0}
        \setcounter{theorem}{0}
        \setcounter{corollary}{0}
        \setcounter{definition}{0}
        \setcounter{equation}{0}
        \renewcommand{\thefigure}{\Alph{appendixc}.\arabic{figure}}
        \renewcommand{\thetable}{\Alph{appendixc}.\arabic{table}}
        \renewcommand{\theappendixc}{\Alph{appendixc}}
        \renewcommand{\thelemma}{\Alph{appendixc}.\arabic{lemma}}
        \renewcommand{\thetheorem}{\Alph{appendixc}.\arabic{theorem}}
        \renewcommand{\thedefinition}{\Alph{appendixc}.\arabic{definition}}
        \renewcommand{\thecorollary}{\Alph{appendixc}.\arabic{corollary}}
        \renewcommand{\theequation}{\Alph{appendixc}.\arabic{equation}}
        \noindent{\tenbf Appendix \theappendixc #1}\par\vspace{5pt}}
\newcommand{\subappendix}[1] {\vspace{12pt}
        \refstepcounter{subappendixc}
        \noindent{\bf Appendix \thesubappendixc. {\kern1pt \bfit #1}}
	\par\vspace{5pt}}
\newcommand{\subsubappendix}[1] {\vspace{12pt}
        \refstepcounter{subsubappendixc}
        \noindent{\rm Appendix \thesubsubappendixc. {\kern1pt \tenit #1}}
	\par\vspace{5pt}}

\newcommand{\textlineskip}{\baselineskip=13pt}
\newcommand{\smalllineskip}{\baselineskip=10pt}

\def\eightcirc{
\begin{picture}(0,0)
\put(4.4,1.8){\circle{6.5}}
\end{picture}}
\def\eightcopyright{\eightcirc\kern2.7pt\hbox{\eightrm c}} 

\newcommand{\copyrightheading}[1]
	{\vspace*{-2.5cm}\smalllineskip{\flushleft
	{\footnotesize Modern Physics Letters A, #1}\\
	{\footnotesize $\eightcopyright$\, World Scientific Publishing
	 Company}\\
	 }}

\newcommand{\publisher}[2]{{\begin{center}\footnotesize\smalllineskip 
	Received #1\\
	Revised #2
	\end{center}
	}}

\def\abstracts#1#2#3{{
	\centering{\begin{minipage}{4.5in}\footnotesize\baselineskip=10pt
	\parindent=0pt #1\par 
	\parindent=15pt #2\par
	\parindent=15pt #3
	\end{minipage}}\par}} 

\newcommand{\bibit}{\nineit}
\newcommand{\bibbf}{\ninebf}
\renewenvironment{thebibliography}[1]
	{\frenchspacing
	 \ninerm\baselineskip=11pt
	 \begin{list}{\arabic{enumi}.}
        {\usecounter{enumi}\setlength{\parsep}{0pt}     
	 \setlength{\leftmargin 12.7pt}{\rightmargin 0pt} 
         \setlength{\itemsep}{0pt} \settowidth
	{\labelwidth}{#1.}\sloppy}}{\end{list}}

\newcounter{itemlistc}
\newcounter{romanlistc}
\newcounter{alphlistc}
\newcounter{arabiclistc}

\newcommand{\fcaption}[1]{
        \refstepcounter{figure}
        \setbox\@tempboxa = \hbox{\footnotesize Fig.~\thefigure. #1}
        \ifdim \wd\@tempboxa > 5in
           {\begin{center}
        \parbox{5in}{\footnotesize\smalllineskip Fig.~\thefigure. #1}
            \end{center}}
        \else
             {\begin{center}
             {\footnotesize Fig.~\thefigure. #1}
              \end{center}}
        \fi}

\newcommand{\tcaption}[1]{
        \refstepcounter{table}
        \setbox\@tempboxa = \hbox{\footnotesize Table~\thetable. #1}
        \ifdim \wd\@tempboxa > 5in
           {\begin{center}
        \parbox{5in}{\footnotesize\smalllineskip Table~\thetable. #1}
            \end{center}}
        \else
             {\begin{center}
             {\footnotesize Table~\thetable. #1}
              \end{center}}
        \fi}

\def\@citex[#1]#2{\if@filesw\immediate\write\@auxout
	{\string\citation{#2}}\fi
\def\@citea{}\@cite{\@for\@citeb:=#2\do
	{\@citea\def\@citea{,}\@ifundefined
	{b@\@citeb}{{\bf ?}\@warning
	{Citation `\@citeb' on page \thepage \space undefined}}
	{\csname b@\@citeb\endcsname}}}{#1}}

\newif\if@cghi
\def\cite{\@cghitrue\@ifnextchar [{\@tempswatrue
	\@citex}{\@tempswafalse\@citex[]}}
\def\citelow{\@cghifalse\@ifnextchar [{\@tempswatrue
	\@citex}{\@tempswafalse\@citex[]}}
\def\@cite#1#2{{$\null^{#1}$\if@tempswa\typeout
	{IJCGA warning: optional citation argument 
	ignored: `#2'} \fi}}

\def\pmb#1{\setbox0=\hbox{#1}
	\kern-.025em\copy0\kern-\wd0
	\kern.05em\copy0\kern-\wd0
	\kern-.025em\raise.0433em\box0}

\def\fnt#1#2{\footnotetext{\kern-.3em
	{$^{\mbox{\scriptsize #1}}$}{#2}}}

\def\fpage#1{\begingroup
\voffset=.3in
\thispagestyle{empty}\begin{table}[b]\centerline{\footnotesize #1}
	\end{table}\endgroup}

\def\runninghead#1#2{\pagestyle{myheadings}
\markboth{{\protect\footnotesize\it{\quad #1}}\hfill}
{\hfill{\protect\footnotesize\it{#2\quad}}}}
\font\tenrm=cmr10
\font\tenit=cmti10 
\font\tenbf=cmbx10
\font\bfit=cmbxti10 at 10pt
\font\ninerm=cmr9
\font\nineit=cmti9
\font\ninebf=cmbx9
\font\eightrm=cmr8

\def\qed{\hbox{${\vcenter{\vbox{			
   \hrule height 0.4pt\hbox{\vrule width 0.4pt height 6pt
   \kern5pt\vrule width 0.4pt}\hrule height 0.4pt}}}$}}

\renewcommand{\thefootnote}{\fnsymbol{footnote}}	

\begin{document}
\setlength{\textheight}{7.7truein}  

\runninghead{Quantum gravity corrections to 
particle $\ldots$}{Quantum gravity corrections to 
particle $\ldots$}

\normalsize\textlineskip
\thispagestyle{empty}
\setcounter{page}{1}

\copyrightheading{}			

\vspace*{0.88truein}

\fpage{1}
\centerline{\bf QUANTUM GRAVITY CORRECTIONS TO}
\baselineskip=13pt
\centerline{\bf PARTICLE INTERACTIONS
}
\vspace*{0.37truein}
\centerline{\footnotesize LUIS  URRUTIA
}
\baselineskip=12pt
\centerline{\footnotesize\it Departamento de F\'\i sica de Altas Energ\'\i as, Universidad Nacional Aut\'onoma de M\'exico. }
\baselineskip=10pt
\centerline{\footnotesize\it Circuito Exterior, C.U., M\'exico, D.F 05410,
M\'exico
}

\vspace*{0.225truein}

\publisher{(received date)}{(revised date)}

\vspace*{0.21truein}
\abstracts{ An heuristic semiclassical procedure that   
incorporates quantum gravity induced corrections in the description of photons and spin 1/2 fermions is reviewed. Such  modifications are calculated in the framework of loop quantum gravity and they arise from the granular structure of space at short distances. The resulting effective theories are described by  power counting non-renormalizable actions which exhibit Lorentz violations at Planck length scale. The modified Maxwell and Dirac equations lead to corrections of the  energy momentum relations for the corresponding particle at such scale. 
An action for the relativistic point particle exhibiting such modified dispersion relations is constructed and the first steps towards the study of a consistent coupling between these effective theories are presented.}{}{}



\vspace*{1pt}\textlineskip	
\section{Introduction }	
\vspace*{-0.5pt}
\noindent
Recently there has been a revival in the  interest of  studying  both  observational and theoretical  manifestations of quantum gravity induced effects.\cite{0,1,1.5,2} This means to consider phenomena associated with the Planck scale, which are highly suppressed in standard scenarios. Many of the envisioned effective theories for particles which incorporate  Planck  scale modifications also induce minute violations of Lorentz covariance at such  scale. In this way, these studies naturally overlap with the systematic approach developed by Colladay and  Kostelecky \cite{3} which provides the most general power counting renormalizable extension of the standard model that incorporates both Lorentz and CPT violations. This framework has been used to set experimental bounds upon the interactions that produce such violations and the observations performed so far cover a wide range  of experimental settings.\cite{4} As pointed out some time ago, the propagation of high energy particles through cosmological distances may provide a realistic possibility to observe one of the effects
associated with  Planck scale corrections: the modification of the  corresponding energy-momentum relations.\cite{1} In the case of photons the authors of Ref.(1)  proposed to consider the modified dispersion relations
\begin{equation}
\label{MDR}
c^2 \, {\vec p}{\,\,}^2=E^2(1 + \xi \, E/E_{QG}+ O(E/E_{QG})^2),  
\end{equation}
with $E_{QG}$ being a scale expected to be close to the Planck mass $M_P=1/\ell_P =10^{19} \, GeV $. The relation  (\ref{MDR}) implies an energy dependent photon velocity leading to a time retardation between two photons  simultaneously emitted  which are  detected with a difference in  energies  $\Delta E$ after traveling a distance $L$. The uncorrected expression for such retardation is  
\begin{eqnarray}
\label{RET}
\Delta t &\approx& \xi \, ({\Delta E}/E_{QG})\,({L}/{c}). 
\end{eqnarray}
In the process of detecting photons from the active galaxy Markarian 421 ($L= 3.5 \,\, l.y.$), Biller et. al.\cite{5} have identified events with $\Delta E= 1 \,TeV$ arriving  to earth within the time resolution of the measurement: $\Delta t= 280 s$. In this way the lower bound ${E_{QG}}/\xi= 4 \times 10^{16}\, GeV $  is established. Time resolutions of milliseconds have been achieved in recent
gamma ray burst (GRB) observations\cite{6} and they will substantially improve up to $10^{-7} s$ with the Gamma Ray Large Area Telescope
(GLAST) to be operating in the International Space station by 2006. Nevertheless, the cosmic radiation background (CRB) prevents  space to be transparent to the propagation of 
very high energy photons. For this reason, the detection of ultrahigh energy neutrinos (UHEN) could provide an arena  to observe such effects. In fact, the fireball model for the emission of GRB predicts also the generation of $10^{14}-10^{19} eV$ neutrino bursts. The Extreme Universe Space Observatory (EUSO), already approved for accommodation study on  the International Space Station would measure such UHEN, also in coincidence with the photons associated to the burst. As we can see, the near future might provide us with very good chances to detect such quantum gravity effects, or at least to set rather strict bounds on the theories predicting them. As a matter of fact,  constraints upon the parameters defining such Lorentz violating theories  have already been established by using current observations\cite{7}. The extension of the modified dispersion relations (\ref{MDR}) to other particles has provided a way  to circumvent some 
traditional astrophysical paradoxes such as the existence of the GZK cutoff, the observation of multi-TeV photons from Markarian 501 and the pion-stability  paradox related to the structure of the air showers  produced by high energy  cosmic rays.\cite{8} The pioneering work revealing the appearance of Planck scale modifications to photon propagation in the framework of loop quantum gravity \cite{10} was made by  Gambini and Pullin.\cite{9} 
Also in the framework of loop quantum gravity the present author, in a   collaboration with  J. Alfaro  and   H. Morales-T\'ecotl, has extended such results by developing an heuristic semiclassical approach which allows the construction of effective field theories for photons and neutrinos, including Planck scale corrections arising as a manifestation of quantum gravitational effects.\cite{11} In particular, dispersion relations of the form (\ref{MDR}) arise from such construction.
An alternative approach inspired in string theory \cite{11.5} has been  developed by Ellis et. al. and includes
both photons and spin $1/2$ particles.\cite{12} 
The paper is organized as follows: in section 2
a brief review of the basic assumptions underlying the semiclassical approximation developed in Ref.(11) is given. The remaining sections contain new material which can be considered as the first steps to investigate a fully consistent
coupling of the effective theories involved. Section 3 contains the formulation of the Gambini-Pullin electrodynamics with sources in terms of the standard electromagnetic 
potentials. In section 4, an action for the relativistic particle yielding Planck scale modified dispersion relations is presented. Finally, in section 5,  I discuss  the extension to a Dirac particle of the effective action previously found for two-components spin 1/2 fermions.

\setcounter{footnote}{0}
\renewcommand{\thefootnote}{\alph{footnote}}
\section{The Effective Theories}
\noindent
Each effective matter Hamiltonian is defined as the expectation value of the corresponding quantum gravity operator in a semiclassical mixed state which describes a flat metric together with the corresponding matter field. The requirements and properties of such a state are made precise in the sequel. The resulting effective theories violate Lorentz covariance at
the Planck scale and such violation can be understood as a spontaneous symmetry breaking generated when taking the  expectation value in the semiclassical state.

In this section I  summarize the procedure for the case of (the magnetic sector of)  Maxwell theory. The starting point is the corresponding Hamiltonian
\begin{equation}
H = \int_{\Sigma} d^3x \,\frac{1}{2}\,\frac{q_{ab}}{\sqrt{\det q}}
[\underline E ^a \underline E^b + \underline B ^a \underline B^b],
\label{MHAM}
\end{equation}
where  the space time is assumed to be a manifold $M$ with  topology $\Sigma\times I\!\!R$. Here $\Sigma$ is a Riemannian 3-manifold with metric  $q_{ab}$, ${\underline E}^a $ and  ${\underline B}^a$ denote the electric and magnetic fields respectively. Thiemann has proposed a general regularization scheme that produces a sound mathematical definition for all the operators entering in the  description of loop  quantum gravity.\cite{13} Such regularized operators act upon states which are  functions of generalized connections defined over graphs. A basis for such space is provided by the so called spin network states. A graph $\Gamma $ is a set of vertices $v\in V(\Gamma)$ in $\Sigma$ which are joined by edges $e$.  The regularization procedure is based upon a triangulation of space which is adapted to each graph. This means that the space surrounding any vertex of $\Gamma$ is filled with   tetrahedra
$\Delta $ having only one vertex in common with  the graph (called the basepoint $v(\Delta)$) plus segments $s_I(\Delta)$ starting at $\Delta $  and directed along   the edges of the graph. In the regions not including the vertices of $\Gamma$  the choice of tetrahedra is arbitrary and the results are independent of it.  The arcs connecting the end points of $s_I(\Delta)$ and $s_J(\Delta)$  are denoted by $a_{IJ}(\Delta)$ and the loop $\alpha_{IJ}:= s_I\circ
a_{IJ}\circ s_J^{-1}$ can be formed. A fundamental property of  this procedure is the use of the volume operator ${\hat V}$ as a convenient regulator. In this way, the action of the operators is finite and gets concentrated only in the vertices of the graph. 

In the case of the magnetic sector of (\ref{MHAM}) Thiemann's regularization leads to the operator
\begin{eqnarray}
{\hat H}^{B} &=&\frac{1}{2\,\hbar ^{2}\kappa ^{2}}\sum_{v\in V(\Gamma
)}\,\left( \frac{2}{3!}\frac{8}{E(v)}\right) ^{2}\sum_{v(\Delta )=v({%
\Delta ^{\prime }})=v}\epsilon ^{JKL}\,\epsilon ^{MNP}\times  \nonumber \\
&\times &{\hat w}_{i \, L \Delta } \, {\hat w}_{i \, P \Delta' } \,\left( {%
\underline{h}}_{\alpha _{JK}(\Delta )}-1\right)\left(
{\underline{h}}_{\alpha _{MN}(\Delta ^{\prime })}-1\right). \, \label{hm1}
\label{REGHAM}
\end{eqnarray}%
where
\begin{eqnarray}
{\hat w}_{i \, L \Delta } &=& tr \left(\tau_i \, h_{s_L(\Delta)}
\left[h_{s_L(\Delta)}^{-1}, \sqrt{\hat{V}_v}  \right] \right) \label{WILD}
\end{eqnarray}
and $E(v)$ is  a factor related to the number of edges that joint at the vertex $v$. In Eqs.(\ref{REGHAM}, \ref{WILD}) $h_{\alpha(\Delta)} \,\, ({\underline h}_{\alpha(\Delta)})$  denote  parallel transport operators (holonomies)  of either the gravitational connection $A_{ia}$ or the electromagnetic connection ${\underline{A}}_a$, respectively,  along  the path $\alpha$ associated to the tetrahedron $\Delta$. Such holonomies are $SU(2)$ and $U(1)$ group elements, respectively. The corresponding trajectories have been previously defined. 

Next, the  quantum state that produces the semiclassical approximation is described. To this end let us  consider an ensemble of graphs together with their adapted triangulation ( which means a set of segments $\{ s_I(\Delta) \}$ for each graph), characterized by some probability distribution $P(\Gamma)$. To each graph $\Gamma$ one  associates a wave function $|\Gamma, {\cal L}, \underline{{\vec{E}}}, \underline{{\vec{B}}} \rangle $ which is peaked with respect to  the classical electromagnetic field configuration together with a flat gravitational metric and a zero value for the gravitational connection. In other words, the contribution for each operator inside the expectation value  can  be  estimated  as 
\begin{eqnarray}
\langle\Gamma, {\cal L}, \underline{{\vec{E}}}, \underline{{\vec{B}}}|\,
...{\hat q}_{ab}...\,|\Gamma, {\cal L}, \underline{{\vec{E}}}, \underline{{\vec{B}}}\rangle&=&
\delta_{ab} + O\left(\frac{\ell_P}{\cal L}\right)
\nonumber \\
\langle\Gamma, {\cal L}, \underline{{\vec{E}}}, \underline{{\vec{B}}}|\,
...{\hat A}_{ia}...\,|\Gamma, {\cal L}, \underline{{\vec{E}}}, \underline{{\vec{B}}}\rangle&=& 0\, + \frac{1}{\cal L}\, \left(\frac{\ell_P}{{\cal L}}\right)^\Upsilon,
\label{EXPV}
\end{eqnarray}
while the  expectation values including  the electric and magnetic operators are estimated through their corresponding classical values ${\vec E}$ and ${\vec B}$. Not surprisingly, the semiclassical state specifies both the classical coordinate and the classical momentum for each pair of canonical variables. The scale ${\cal L}>> \ell_P$ of the wave function is such that the continuous flat metric approximation is appropriate for distances much larger that ${\cal L}$, while the granular structure of spacetime   becomes relevant when probing distances smaller that ${\cal L}$. Such scale will have a natural realization according to each particular physical situation. In the sequel we set $\Upsilon=0$ for simplicity  and denote the  semiclassical state as $ |\Gamma, S \rangle=
|\Gamma, {\cal L}, \underline{{\vec{E}}}, \underline{{\vec{B}}}\rangle$.  

In a very schematic way I  summarize now the method of calculation. 
For each graph $\Gamma$  the effective Hamiltonian  is defined as ${\rm H}_\Gamma= \langle \Gamma ,S | {\hat H}_{\Gamma} |\Gamma, S \rangle$. For a given vextex, inside the expectation value, one expands each operator in powers of the segments $s_I(\Delta)$ plus derivatives of the matter operators. In the case of (\ref{REGHAM}) this produces
\begin{eqnarray}
{\rm H}^B_{\Gamma} &=&\sum_{v\in V(\Gamma
)}\,\sum_{v(\Delta )=v}\, \langle \Gamma ,S |{\hat {\underline F}}_{p_1q_1}(v)...\partial^{a_1}... {\hat {\underline F}}_{pq}(v){\hat T}_{a_1}...^{pq\,p_1q_1\,...}(v, s(\Delta))|\Gamma, S\rangle. \nonumber\\
\end{eqnarray}
where ${\hat T}$ contains  gravitational operators together with
contributions depending on the segments of the adapted triangulation in
the  particular graph. Next,  space  is considered to be divided into  boxes, each centered
at a given  point ${\vec{x}}$ and with volume ${\cal L}^{3}\approx d^{3}\,x$. The choice of boxes is the same for all the graphs considered.  Each box
contains a large number of vertices of the semiclassical state (${\cal L}>\!>\ell
_{P}$), but it is considered as infinitesimal in the scale where the space can be regarded
as continuous. The sum over the vertices in (\ref{REGHAM}) is subsequently split as the sum over the vertices in each box, plus the sum over boxes. Also, one assumes that the electromagnetic operators are slowly varying within 
a  box (${\cal L}<\!<\lambda $, with $\lambda$ been the photon wavelength), in such a way that for all the vertices inside
a given  box one can write
$\langle \Gamma, S|\dots \underline{{%
\hat{F}}}_{ab}(v)\dots |\Gamma, S
\rangle = \mu {\underline F}_{ab}({\vec{x}}).
$
Here ${\ {\underline F}}_{ab}$ is the classical electromagnetic field at
the center of the box and $\mu$ is a dimensionless constant which is 
determined in such a way that  the standard classical result in the zeroth order approximation is recovered.
Applying  the procedure just described to (\ref{REGHAM}) leads to 
\begin{eqnarray}
{\rm H}_\Gamma^{B}&=&\sum_{{\rm Box}}\,\,\underline{{F}}_{p_{1}\,q_{1}}({\vec{x}}%
)\dots \,\,\left( \partial
^{a_{1}}\dots \,\,\underline{{\ F}}_{p\,q}({\vec{x}}%
)\right) \,\,\sum_{v\in {\rm Box}}\ell _{P}^{3}\nonumber \times\\
&& \sum_{{v(\Delta )=v}}\,\mu ^{n+1} \langle \Gamma, S|\frac{1}{\ell _{P}^{3}}{\hat{T}%
}_{a_{1}\dots }{}^{\,pqp_{1}\,q_{1}\dots }(v,s(\Delta ))|\Gamma, S \rangle.
\label{HAMG}
\end{eqnarray}
In the above, $n+1$ is the total number of factors $F_{pq}({\vec x})$ . The expectation value of the gravitational contribution is expected to be a rapidly varying function inside each box. Finally, the  effective
Hamiltonian is defined as an  average over the graphs $\Gamma$, i.e. over adapted triangulations : ${\rm H}^B=\sum_{\Gamma} P(\Gamma)\, {\rm H}_\Gamma^B $. This  effectively amounts to average the expectation values remaining  in each box of the sum 
(\ref{HAMG}). We call this average ${{T}%
}_{a_{1}\dots }{}^{\,pqp_{1}\,q_{1}\dots }({\vec x})$ and estimate it by demanding $T$ to be constructed from the  flat space tensors
$\delta_{ab}$ and $ \epsilon_{abc}$. In this way one is  imposing isotropy and rotational invariance on our final result.  Also  the scalings given in  (\ref{EXPV}) together with the additional assumptions: $\langle \Gamma, S|...{\hat V}...|\Gamma, S\rangle \longrightarrow \ell_P^3,  s_I^a \longrightarrow \ell_P$ are used . Let us remark that the above average can be understood as taking the expectation value of the Hamiltonian (\ref{REGHAM}) in a mixed state characterized by the density matrix $\rho=\sum_{\Gamma} |\Gamma, S\rangle\, P(\Gamma)\, \langle \Gamma, S|$. After replacing the summation over boxes by the integral over space, the resulting Hamiltonian has the final form
\begin{equation}
{\rm H}^{B} =\int d^{3}x\ \underline{{\ F}}_{p_{1}\,q_{1}}({%
\vec{x}})\dots \,\left( \frac{{}%
}{{}}\partial ^{a_{1}}\dots \underline{{\ F}}_{pq}({\vec{x}}%
)\right) \,\,{{T}}_{a_{1}\dots }{}^{\,pqp_{1}\,q_{1}\dots}({\vec{x}}).  \label{AMH}
\end{equation}%

Some comments are now in order. A rigorous semiclassical treatment of loop quantum gravity is still in the process of development.\cite{14} Since the  approach presented here has made use only of the main features that semiclassical states should have, all dimensionless coefficients in the expectation values that contribute to ${{T}}_{a_{1}\dots }{}^{\,pqp_{1}\,q_{1}\dots}({\vec{x}})$ in (\ref{AMH}) remain
undetermined. Besides, the calculation has not been performed in a covariant way. On the contrary, the results are expected to be valid only in a preferred reference frame. Thus, the states  considered so far will not annhilate the Hamiltonian constraint of quantum gravity.

By applying the above procedure to photons ($\gamma$) and to two-component massive neutrinos with definite chirality ($\nu$) the following
results, including corrections to order $\ell_P^2$, are obtained \cite{11}
\begin{eqnarray}
{\rm H}^{\gamma}= \int d^3{\vec x} \left[\left(1+ \theta_7
\,\left(\frac{\ell_P}{{\cal L}}\right)^{2} \right)\frac{1}{2}\left(\frac{}{} 
\underline{{\vec B}}^2 + \underline{{\vec E}}^2\right)
+ \theta_8 \ell_P \left( \frac{}{} \underline{\vec B}%
\cdot(\vec\nabla \times\underline{\vec B})+ \underline{\vec E}%
\cdot(\vec\nabla \times\underline{\vec E}) \right)\right. && \nonumber \\
\left.+ \theta_3 \, \ell_P^2 \, \left( \frac{}{}\underline{B}%
^a \,\nabla^2 \underline{B}_a + \underline{E}^a \,\nabla^2 \underline{E}%
_a\right)+ \theta_2\,\ell_P^2\,{\underline E}^a \partial_a \partial_b {%
\underline E}^b  + \theta_4\, {\cal L}^2 \, \ell_P^2 \, \, \left(\frac{}{}%
\underline{{\vec B}}^2\right)^2 +\dots \right]. && \nonumber\\
\label{PHOTEFF}
\end{eqnarray}
\begin{eqnarray}  
{\rm H}^\nu &=& \int d^3 x \left[ i \
\pi(\vec x) \left(1 + {\kappa}_{1} \frac{\ell_P}{%
{\cal L}} +... -%
\kappa_2 \ \ell_P^2 \ \nabla^2 \right)\tau^d\partial_d \xi({\vec x}) \, + \right.\nonumber\\
 &+& \left.\frac{i}{{\cal L}} \ \pi({\vec x}) \left( {%
\kappa}_{3} \frac{\ell_P}{{\cal L}}+... - \kappa_{4} \, \ell_P^2
\nabla^2 \right) \xi({\vec x}) \right. \nonumber \\
 &+& \left. \frac{m}{2 } \xi^T({\vec x})\ (i \sigma^2)\left( 1 + { \kappa}%
_{5} \frac{\ell_P}{{\cal L}} + { \kappa}_{6} \ \ell_P \ \tau^a
\partial_a \right)\xi({\vec x}) + c.c.  \right].
\label{EFNF}
\end{eqnarray}
Here $\pi({\vec x})=i \xi^*({\vec x})$ and $\tau^a=-(i/2)\,\sigma^a\, $ with $\sigma^a$ being the standard Pauli matrices. In particular, the effective theories given in (\ref{PHOTEFF}) and  (\ref{EFNF}) imply Lorentz violating Planck scale modifications of the corresponding particle energy-momentum relations which are calculated in Ref.(11). In both cases the scale ${\cal L}$ has been  estimated as  the De Broglie wavelength of the corresponding particle. In the case of neutrinos the condition $m_\nu << p_\nu$ was assumed.

Before closing this section let us emphasize that the effective theories (\ref{PHOTEFF}) and (\ref{EFNF}) are expected to be valid in a particular reference frame, the most natural one being that in which the CRB spectrum looks isotropic. This means that the involved  scales $\ell_P$ and ${\cal L}$ will experience FitzGelrald-Lorentz contraction in going to the laboratory frame, for example. Also, the velocity of light will not have a universal value, exhibiting corrections  depending on $\ell_P$ which arise from the modified dispersion relations. An alternative  point of view allowing for deformed dispersion relations which are valid in every reference frame has been recently proposed\cite {14.7} and further elaborated.\cite{15} This requires the formulation of a relativity principle having two observer independent scales: the speed of light constant and the Planck-length constant, which can be realized via non-linear realizations of the Lorentz group. An analysis of the common features and main differences between  the  approaches of Refs.(19) and (20) has also appeared.\cite{15.5}

\setcounter{footnote}{0}
\renewcommand{\thefootnote}{\alph{footnote}}

\section{The Gambini-Pullin Electrodynamics with Sources}
\noindent
In this section I present the first steps towards  a more detailed discussion of the modified electrodynamics obtained  in (\ref{PHOTEFF}). In order to have the correct normalization in the zeroth order case ($\ell_P=0$) 
it is convenient to  make the field redefinition
\begin{eqnarray}
( 1+\theta _{7}\,( \ell _{P}/{\cal L}) ^{2})^{1/2} {\underline E}_i &\longrightarrow &E_i  
\end{eqnarray}%
and similarly for ${\underline B}_i \longrightarrow B_i$.
Considering only the contribution linear in $\ell_P$ one is  left with  the effective Hamiltonian density 
\begin{eqnarray} 
{\cal H}^{EM}=\frac{1}{2}\left( \vec{B}%
^{2}+\vec{E}^{2}\right) +\theta _{8}\ell _{P}\left( \vec{B}%
\cdot (\vec{\nabla}\times \vec{B})+\vec{E}\cdot (\vec{\nabla}\times \vec{E}%
)\right),
\label{HEM}
\end{eqnarray}
which was previously obtained by Gambini and Pullin.\cite{9} This theory predicts birefringence effects which are manifest through  different propagation velocities for left and right polarized photons. Such velocity difference is proportional to  the parameter  $\theta_8$ in Eq. (\ref{HEM}) and it is linear in  $\ell_P$. By analizing the presence of linear polarization in the optical and ultraviolet spectrum of some cosmological sources the bound $\theta_8 < 10^{-3}$ is obtained.\cite{14.5}

Adding the appropriate sources in  the first order action
\begin{eqnarray}
S[\Phi, A_i, \, E_j]=\int dt\,d^3{\vec x}\left(-E_i\,{\dot A}_i-{\cal H}^{EM}+ \Phi \left(\partial_i\,E_i -4\pi\,\rho \right)+ 4\pi J_iA_i \right),
\end{eqnarray}
with ${\vec B}=\nabla\times {\vec A}$ and the potential $\Phi$ acting as a Lagrange multiplier, the modified Maxwell equations
\begin{eqnarray}
\nabla \cdot \vec{E}=4\pi \rho ,\;\;\;\;\;\nabla \cdot \vec{B},&=&0\nonumber \\
\vec{\nabla}\times\left( \vec{B}+2\theta _{8}\ell _{P}\nabla \times 
\vec{B}\right)-\frac{\partial \vec{E}}{\partial t}%
&=& 4\pi \vec{J}, \nonumber\\
\vec{\nabla}\times \left( \vec{E}+2\theta _{8}\ell _{P} \nabla \times 
\vec{E} \right) + \frac{\partial \vec{B}}{\partial t} &=&0,
\label{MODME}
\end{eqnarray}
are obtained.
From Eqs. (\ref{MODME}) one can prove the continuity equation for the electric charge as a signal of consistency. In order to write a second order Lagrangian formulation of Maxwell equations (\ref{MODME}) in terms of the basic fields $A^\mu=(\Phi, {\vec A})$ it is convenient to reintroduce the potentials starting, as usual, from the homogeneous equations. Here I use the definition
$F_{\mu\nu}=\partial_\mu A_\nu-\partial_\nu A_\mu$
and the conventions of Jackson. The magnetic field retains the standard relation with the vector potential: $B_{i}=\epsilon
_{ijk}F_{jk}$, while the relation defining the electric field is changed to
\begin{eqnarray}
&&\vec{E}+2\theta _{8}\ell _{P}\, \nabla \times \vec{E}=-\nabla
\Phi -\frac{1}{c}\frac{\partial \vec{A}}{\partial t}=F_{0i}.
\label{ECFP}
\end{eqnarray}
 Eq. (\ref{ECFP}) implies that the resulting Lagrangian will be non local. In order to invert (\ref{ECFP}), the operator
\begin{equation}
M _{ij}^{-1}(x,y)=\delta^4(x-y)\left(\delta _{ij}+2\theta _{8}\ell _{P}\epsilon
_{ikj}\partial _{k}\right)=\delta(x^0-y^0)M^{-1}_{ij}({\vec x}, {\vec y}),
\end{equation}
is defined in such a way that
\begin{equation}
E_{i}(x)=\int d^4y\, M_{ij}(x,y)F_{0j}(y).
\label{ECFF}
\end{equation}
The Lagrangian  density is
${\cal L}=-E^{i}\;\dot{A}_{i}-{\cal H}$,  
where  the velocities ${\dot A}_i$ are introduced  via the equation (\ref{ECFF}), which defines $E_i$ as a non local function of the potentials. This leads to
\begin{equation}
{\cal L}=\;\frac{1}{2}\left( {{\vec{E}}}^{2}-{{\vec{B}}}^{2}\right) +\theta
_{8}\ell _{P}\;\left( {\vec{E}}\cdot (\vec{\nabla}\times {\vec{E}})-{\vec{B}}%
\cdot (\vec{\nabla}\times {\vec{B}})\right)-4\pi\,J^\mu A_\mu.
\label{MLDM} 
\end{equation}
 The above Lagrangian density  violates P, but preserves C and T. The CPT violation is produced because (\ref{MLDM}) is both non local and  non invariant under proper Lorentz transformations.
 Nevertheless, (\ref{MLDM}) is invariant under rotations and thus it is valid in a preferred coordinate system which we identify with that in which  the CRB looks isotropic. It is  interesting to observe that the Lagrangian density (\ref{MLDM}) is  power counting non-renormalizable, thus constituting  a case which is  not considered in the general framework of  Ref.(3) to discuss Lorentz and CPT violations. All effective theories generated via the framework presented here are expected to enjoy such non-renormalizability property. 
Also, (\ref{MLDM}) describes an effective theory which should be valid only at energy scales much lower than the Planck mass. 
A convenient way of presenting the  action arising from (\ref{MLDM}) is 
\begin{equation}
S=\int d^{4}xd^{4}y\;\frac{1}{2}\left( E_{i}(x)\;M_{iq}^{-1}(x,y)\;E_{q}(y) 
-B_{i}(x)\;M_{iq}^{-1}(x,y)\;B_{q}(y)\right)-4\pi\int d^{4}x J^\mu A_\mu 
\label{ENLA}
\end{equation}
where the operator $M_{iq}^{-1}(x,y) $ satisfies the additional symmetry condition
$M_{iq}^{-1}(x,y)\\ =M_{qi}^{-1}(y,x)$. Varying the action (\ref{ENLA}) leads indeed to the  modified Maxwell equations (\ref{MODME}). Before closing  this section I discuss the explicit inversion of the operator $M_{iq}^{-1}({\vec x}, {\vec y}) $. In momentum space, with the convention  $\nabla =i\;\vec{k} $, one has
\begin{equation}
\left( M^{-1}\right) _{im}=\delta _{im}+\left( i\kappa \ell _{P}\right)
\epsilon _{ilm}\;k_{l},\;\;\varsigma = i\kappa \ell _{P}, 
\end{equation}
which inverse is
\begin{equation}
M_{ij}=\frac{1}{1+\varsigma ^{2}\vec{k}^{2}}\left( \delta _{ij}+\varsigma
\epsilon _{ijp}k_{p}+\varsigma ^{2}\;k_{i}k_{j}\right). 
\label{INVER}
\end{equation}
The pole in (\ref{INVER}) signals the need of a  cutoff in order to regulate the Fourier transform. This is in accordance with the effective character of the theory, which is no more valid for momenta close to the pole position
 $k_{\infty}=1/(\kappa \ell _{P})$.
In coordinate representation, (\ref{INVER}) leads to  the Fourier transform
\begin{equation}
M(\vec{z})=\int d^{3}k\;e^{i\vec{k}\cdot \vec{z}}\frac{1}{1-\kappa
^{2}\ell _{P}^{2}\vec{k}^{2}}= \frac{4\pi}{z}\int kdk\;\;\frac{\sin kz}{1-\left( \kappa
^{2}\ell _{P}^{2}\right) k^{2}}.
\label{INCR}
\end{equation}
For energies which are low compared to $k_{\infty}$, it will be enough to consider the local approximation to the operator $M_{ij}({\vec x},{\vec y})$, which  in turn leads to a simpler local effective theory. This is done by expanding the denominator of  (\ref{INCR}) in power series of $\ell_P$ and integrating. The result, up to second order in $\ell_P$ is 
\begin{equation}
M_{ij}(\vec{z})=\left( 2\pi \right) ^{3}\delta ^{3}(\vec{z})
\,\left( \delta _{ij}+\kappa \ell
_{P}\epsilon _{ijp}\partial _{p}+\kappa ^{2}\ell _{P}^{2}\left( \partial
_{i}\partial _{j}-\delta _{ij}\nabla ^{2}\right) +O(\kappa ^{3}\ell
_{P}^{3})\;\right), 
\end{equation}
with ${\vec z}={\vec x}-{\vec y}$.

\section{An Action for the Relativistic Particle}
\noindent
In order to consistently couple  the electrodynamics considered in section 3 to a point particle it is necessary to have an action for the latter which  naturally incorporates corrections to the dispersion relations of the type discussed previously. A simple way to do this is by starting from a first
order action where the corresponding modified energy-momentum  relation is incorporated as a constraint. In other words, suppose that one needs to consider  the modified dispersion relation 
\begin{equation}
F(p^{0})-(\vec{p}^{2}+m_{0}^{2})=0.
\label{CONST}
\end{equation}%
The action to consider is
\begin{equation}
S=\int d\tau \left( p_{0}\;\dot{x}^{0}-p^{i}\;\dot{x}^{i}-\frac{\lambda }{2}%
\left( F(p^{0})-(\vec{p}^{2}+m_{0}^{2})\right) \right), 
\end{equation}
where $\lambda$ is a Lagrange multiplier.
The equations of motion are%
\begin{eqnarray}
\delta p_{0} &:&\;\dot{x}^{0}=\frac{\lambda }{2}\frac{d F}{d
p^{0}}\Longrightarrow \frac{2\dot{x}^{0}}{\lambda }=G(p^{0})\Longrightarrow
p^{0}=G^{-1}\left( \frac{2\dot{x}^{0}}{\lambda }\right), \nonumber \\
\delta p^{i} &:&-{\dot x}_i+\lambda p^{i}=0\Longrightarrow p^{i}=\frac{\dot{x}^{i}}{%
\lambda },
\end{eqnarray}%
together with (\ref{CONST}). Here $G(u)=dF(u)/du$. 
Substituting $p^0$ and $p^i$ back into the action yields
\begin{eqnarray}
S &=&\int d\tau \left( \;\dot{x}^{0}G^{-1}\left( \frac{2\dot{x}^{0}}{\lambda 
}\right) -\frac{\lambda }{2}F\left(G^{-1}\left( \frac{2\dot{x}^{0}}{\lambda }%
\right) \right)-\;\frac{\dot{x}^{i}\dot{x}^{i}}{2\lambda }+\frac{\lambda }{2}%
m_{0}^{2}\right).
\label{PFOA}
\end{eqnarray}
The second order action is obtained by eliminating  the auxiliary field $\lambda$  via its  equation of motion arising  from (\ref{PFOA}). In order to illustrate the procedure in a more tractable  situation I choose
\begin{eqnarray}
F(p^{0}) &=&\left( p^{0}\right) ^{2}+\alpha \ell _{P}\frac{1}{3}\left(
p^{0}\right) ^{3},\qquad 
G(p^{0}) =2p^{0}+\alpha \ell _{P}{p^{0}}^{2}=y,\;\; \nonumber \\
G^{-1}\;(y) &=&\frac{1}{\alpha \ell _{P}}\left( \sqrt{\left( 1+\alpha \ell
_{P}y\right) }-1\right).
\label{CUCHO}
\end{eqnarray}
In the approximation linear in $\ell_P$ I obtain
\begin{equation}
G^{-1}\;(y)=\frac{1}{2}y-\frac{1}{8}\alpha \ell _{P}y^{2}, \quad   
F(G^{-1}(y))=\frac{1}{4}y^{2}-\frac{1}{12}\alpha \ell _{P}y^{3}.
\end{equation}%
Clearly, the dispersion relation arising from the choice (\ref{CUCHO}) together with (\ref{CONST}) reproduces  Eq.(\ref{MDR}) in the zero mass limit.
Substituting in (\ref{PFOA}), and after some algebra,  one obtains
\begin{eqnarray}
S &=&\int d\tau \left[ \frac{v^{2}}{2\lambda }-\frac{1}{6}\alpha \ell _{P}%
\frac{\left( \dot{x}^{0}\right) ^{3}}{\lambda ^{2}}+\frac{\lambda }{2}%
m_{0}^{2}\right] ,\;\;v^{2}=\left( \dot{x}^{0}\right) ^{2}-\left( \dot{x}%
^{i}\right) ^{2}.
\end{eqnarray}
The equation of motion for $\lambda$ is 
\begin{eqnarray}
-\frac{v^{2}}{\lambda ^{2}}+\frac{2}{3}\alpha \ell _{P}\frac{\left( \dot{x}%
^{0}\right) ^{3}}{\lambda ^{3}}+m_{0}^{2} &=&0,
\end{eqnarray}%
which is also solved to first order in $\ell_P$ by making the ansatz 
$\lambda =\lambda _{0}\left( 1+\ell _{P}\lambda _{1}\right)$. The result is
\begin{equation}
\label{LAMD}
 \lambda _{0}=\frac{%
\sqrt{v^{2}}}{m_{0}}, \qquad \lambda _{1} =-\frac{1}{3}\alpha \frac{m_{0}\left( \dot{x}^{0}\right) ^{3}%
}{\left( v^{2}\right) ^{3/2}}, 
\end{equation}%
leading to the action 
\begin{equation}
S=m_{0}\int dt\;\left[ \sqrt{1-\vec{v}^{2}}-\frac{\alpha }{6}\left(
m_{0}\ell _{P}\right) \frac{1}{\left( 1-\vec{v}^{2}\right) }\right].
\label{MPACT}
\end{equation}%
Observe that the limit $\ell_P=0$ correctly reproduces the well known relativistic action for the point particle. From the previous equations one can write the energy and momentum in terms of the velocity as 
\begin{equation}
p^{0}=m_{0}\gamma -\frac{\alpha }{2}\ell _{P}m_{0}^{2}\gamma ^{2}\left( 1-%
\frac{2}{3}\gamma ^{2}\right),
\quad
p^{i}=m_{0}\gamma v^{i}\left( 1+\frac{\alpha }{3}\left( m_{0}\ell
_{P}\right) \gamma ^{3}\right).
\label{RES}
\end{equation}%
At this stage it is appropriate to compare the results  (\ref{RES}) with those of Ref.(20), which, 
to first order in $\ell_P$,  are
\begin{equation}
p^{0}=\;m_{0}\gamma -\ell _{P}m_{0}^{2}\gamma ^{2}, \qquad
p^{i} =m_{0}\gamma v^{i}\left( 1-\ell _{P}m_{0}\gamma \right).
\end{equation}%
The discrepancy is telling us that  the action (\ref{MPACT}) is not an scalar under the specific non-linear representation of the Lorentz group proposed in Ref.(20). Finally, it  is interesting to emphasize that Planck scale corrections to either particle propagation or
quantum field interactions need not necessarily imply violations of Lorentz covariance.\cite{14.7,15,15.5,16}

\section{A  Lagrangian for Dirac Particles}
\noindent
The aim in this section is to construct a modified Lagrangian density for Dirac particles, starting from the theory given by (\ref{EFNF}).  Such two component theory can be  embedded into a four component realization by demanding $\Psi$ to be a Majorana spinor
\begin{equation}
\Psi=\left[ 
\begin{array}{c}
\xi  \\ 
\chi %
\end{array}%
\right]
 =\left[ 
\begin{array}{c}
\xi \\ 
-i\ \sigma^{2}\xi ^{\ast }%
\end{array}%
\right].
 \end{equation}
In this notation, the equations arising from (\ref{EFNF}) are
\begin{eqnarray}
&&\left[ i \frac{\partial }{\partial t}-i{\hat{A}}\ {\vec{%
\sigma}}\cdot \nabla \right] \xi 
-m\left(1 -i C\,\ell_P\,{\vec{\sigma}}\cdot\nabla \right) \chi
=0, \nonumber \\
&&\left[ i\frac{\partial }{\partial t}+i{\hat{A}}\ {\vec{%
\sigma}}\cdot \nabla \right] \chi 
- m\left( 1 -i C\,\ell_P\,{\vec{\sigma}}\cdot\nabla \right) \xi=0,
\label{MAJEQ}
\end{eqnarray}%
with ${\hat{A}}=\left( 1+D\, \ell _{P}^{2}\, \nabla ^{2}\right)$ and  $C,D$ being constants. For simplicity I discuss  here the rather unrealistic  case  ${\cal L}\rightarrow \infty$. Next,  $\chi$ and $\xi$ are considered to be  independent spinors. In the conventions where $\gamma_5$ is diagonal,  $(\gamma_0)^2=1$ and the signature  is $(+---)$, one can  verify that
\begin{equation}
\label{DIRAC}
\left( i\gamma ^{\mu }\partial _{\mu }+i\left(D\,\ell _{P}^{2}\,
\nabla ^{2}\right) \;\vec{\gamma}\cdot \nabla\, -m\left( 1-iC\ell_P\, {\vec \Sigma}\cdot \nabla\right) \right) \Psi =0
\end{equation}
reproduces the equations (\ref{MAJEQ}). The spin operator is given by $ \Sigma^k=(i/2) 
\epsilon _{klm}\gamma ^{l}\gamma ^{m}$.
The Lagrangian density that yields Eq. (\ref{DIRAC}) is
\begin{equation}
{\cal L}_{D} =\frac{1}{2}\left(\frac{}{} i\bar{\Psi}\gamma ^{\mu }\left( \partial _{\mu
}\Psi \right) 
 +iD \, \ell _{P}^{2} \left( \nabla ^{2}\bar{\Psi}%
\right) \gamma ^{k}\left( \partial _{k}\Psi \right)
-m{\bar \Psi}\,(1-iC\ell_P\Sigma^k\,\partial_k )\Psi + h.c. \right),
\label{DIRLAG} 
\end{equation}
which  is invariant under the  global phase transformation
$\delta \Psi =i\delta \Theta \Psi $.
The associated Noether current, which  will be  identified with the electromagnetic current is  
\begin{equation}
J^{0} =\bar{\Psi}\gamma ^{0}\Psi,  \qquad 
J^{k} =\bar{\Psi}(\gamma ^{k}+C\, m\ell_P\, \Sigma^k)\Psi    ,
\end{equation}
to first order in $\ell_P$. The coupling with the modified electrodynamics (\ref{MLDM}) is made in the standard gauge invariant way  via the replacement $\partial_\mu \rightarrow \partial_\mu + ie A_\mu$ in (\ref{DIRLAG}).

\nonumsection{Acknowledgments}
\noindent
The author wishes to thank D.V. Ahluwalia and
N. Dadhich for the warm hospitality  extended to him in Pune. Also he acknowledges H. Morales-T\'ecotl
for many illuminating discussions about loop quantum gravity. This work was supported  in part by grants CONACyT 32431-E and DGAPA IN11700.  

\nonumsection{References}


\noindent

\begin{thebibliography}{000}
\bibitem{0} P. Huet and M. Peskin, {\bibit Nucl. Phys.} {\bibbf B434}, 3 (1995);
J. Ellis, J. L\'opez, N.E. Mavromatos and D.V. Nanopoulos, 
{\bibit Phys. Rev.} {\bibbf D53}, 3846 (1996).
\bibitem{1} 
G. Amelino-Camelia, J. Ellis, N.E. Mavromatos, D.V. Nanopoulos
and S. Sarkar, {\bibit
Nature}\, {\bibbf 393}, 763 (1998). See also the contribution of S. Sarkar to these Procceedings.
\bibitem{1.5}G. Amelino-Camelia, {\bibit Nature} {\bibbf 398}, 216 (1999); {\bibit Lect. Notes Phys.} {\bibbf 541}, 1 (2000).  

\bibitem{2} D.V. Ahluwalia, {\bibit Nature} {\bibbf 398}, 199 (1999);
 G.Z. Adunas, E. Rodriguez-Milla and D.V. Ahluwalia,
{\bibit Phys. Letts.} {\bibbf B485}, 215 (2000); ibid. {\bibit Gen. Rel. Grav.} {\bibbf
33}, 183 (2001).
\bibitem{3}
D. Colladay and V.A. Kostelecky, {\bibit Phys. Rev.} {\bibbf D55}, 6760 (1997), {\bibit Phys. Rev.} {\bibbf D58}, 116002 (1998);V.A. Kostelecky and C.D. Lane,
{\bibit J. Math. Phys.} {\bibbf 40}, 6245 (1999);V.A. Kostelecky and R. Lehnert, {\bibit Phys. Rev.} {\bibbf D63},  065008 (2001); V.A. Kostelecky, {\bibit arXiv: hep-ph/0104227} and references therein; D. Colladay and P. McDonald, {\bibit arXiv:
hep-ph/0202066}. 
\bibitem{4} 
R. Bluhm, {\bibit arXiv: hep-ph/0111323} and references therein.
\bibitem{5} 
S.D. Biller et. al., {\bibit Phys. Rev. Lett.}  {\bibbf 83}, 2108 (1999).
\bibitem{6}
P.N. Bhat et. al., {\bibit Nature} {\bibbf 359}, 217 (1992).

\bibitem{7}V.A. Kostelecky and C.D. Lane, {\bibit Phys. Rev.} {\bibbf D60}, 116010 (1999); J.M. Carmona and J.L. Cort\'es, {\bibit Phys. Letts.} {\bibbf B494}, 75 (2000); C. L\"{a}mmerzhal and C. Bord\'e, in {\bibit 
Lecture Notes in Physics} {\bibbf 562}, Springer 2001 ; R. Brunstein, D. Eischler and S. Foffa, {\bibit arXiv:
 hep-ph/0106309}; S. Liberati, T. Jacobson and D. Mattingly, {\bibit arXiv: hep-ph/0110094}; T. Jacobson, S. Liberati and D. Mattingly, {\bibit arXiv: hep-ph/0112207};  T.J. Konopka and S.A. Major, {\bibit arXiv: hep-ph/0201184};  D. Sudarsky, L. Urrutia and H. Vucetich, {\bibit arXiv: gr-qc/0204027}.
\bibitem{8}
G. Amelino-Camelia and T. Piran, {\bibit Phys. Rev.} {\bibbf D64}, 036005
 (2001); G. Amelino-Camelia, {\bibit Phys. Letts.} {\bibbf B528}, 181 (2002); J. Alfaro and G. Palma, {\bibit arXiv: hep-th/0111176}.
\bibitem{10}
For a review see for example C. Rovelli, {\bibit in  Living Reviews Vol 1}, 1998-1, http://www.livingreviews.org/Articles. See also the contribution of G. Date  to these Procceedings.
\bibitem{9}
R. Gambini and J. Pullin, {\bibit Phys. Rev.} {\bibbf D59}, 124021 (1999). 
\bibitem{11}
J. Alfaro, H. Morales-T\'ecotl and L. Urrutia, {\bibit Phys. Rev. Lett.} 
{\bibbf 84}, 2318 (2000); J. Alfaro, H. Morales-T\'ecotl and L. Urrutia, {\bibit Phys. Rev.} {\bibbf D65}, 103509 (2002); J. Alfaro, H. Morales-T\'ecotl and L. Urrutia in {\bibit JHEP Proceedings, Cartagena de Indias 2000,
High Energy Physics}; J. Alfaro, H. Morales-T\'ecotl and L. Urrutia, in {\bibit  
Proceedings of the Ninth Marcel Grossmann Meeting on General Relativity}, eds. R.T Jantzen, V. Gurzadyan and R. Rufini (World Scientific, 2002).

\bibitem{11.5} V.A. Kostelecky and S. Samuel, {\bibit Phys. Rev.} {\bibbf D39},
683 (1989);  
{\bibit Phys. Rev.} {\bibbf D40}, 1886 (1989); V.A. Kostelecky  and R. Potting,
{\bibit Nucl. Phys.} {\bibbf B 359},545 (1991), {\bibit Phys. Lett.} {\bibbf B 381}, 89 (1996)

\bibitem{12}
J. Ellis, N.E. Mavromatos and D.V. Nanopoulos, {\bibit Gen. Rel. Grav.}
{\bibbf 32}, 127 (2000); J. Ellis, N.E. Mavromatos and D.V.
Nanopoulos, in {\bibit Tegernsee 1999, Beyond the desert}; J. Ellis, K. Farakos, N.E. Mavromatos, V.A. Mitsou and D.V.
Nanopoulos,  {\bibit
Astrophysical Jour.} {\bibbf 535}, 139 (2000);
J. Ellis, N.E. Mavromatos, D.V. Nanopoulos and G.
Volkov, {\bibit Gen.
Rel. Grav.} {\bibbf 32}, 1777 (2000).
\bibitem{13}
T. Thiemann, {\bibit Class. Quan. Grav.} {\bibbf 15}, 1281 (1998);
T. Thiemann, {\bibit Class. Quant. Grav.} {\bibbf 15}, 839 (1998); T. Thiemann, {\bibit Phys. Letts.} {\bibbf B380}, 257 (1996). A complete and self-contained review of these ideas is in  T. Thiemann, {\bibit arXiv: gr-qc/0110034}.
\bibitem{14}
T. Thiemann, {\bibit Class. Quant. Grav.} {\bibbf 18}, 2025 (2001); T. Thiemann and
O. Winkler, {\bibit Class. Quant. Grav.} {\bibbf 18}, 2561 (2001); T. Thiemann and O.
Winkler, {\bibit arXiv: hep-th/0005234}; {\bibit arXiv: hep-th/0005235}; H. Sahlmann and  T. Thiemann, {\bibit arXiv: gr-qc/0102038}.
\bibitem{14.5}
R.J. Gleiser and C.N. Kozameh, {\bibit Phys. Rev.} {\bibbf D64}, 083007 (2001).
\bibitem{14.7} G. Amelino-Camelia, {\bibit Int. J. Mod. Phys.} {\bibbf D11}, 35 (2002).
\bibitem{15}
J. Magueijo and L. Smolin, {\bibit arXiv: hep-th/0112090}.
\bibitem{15.5} G. Amelino-Camelia, D. Benedetti and F. D'Andrea, {\bibit arXiv: hep-th/0201245}. 
\bibitem{16}
T. Padmanabhan, {\bibit Phys. Rev.} {\bibbf 57}, 6206 (1998); K. Srinivasan, L. Sriramkumar and T. Padmanabhan, {\bibit Phys. Rev.} {\bibbf 58}, 044009 (1998);
S. Shankaranarayanan and T. Padmanabhan, {\bibit Int. J. Mod. Phys.} {\bibbf D10}, 351 (2001).

\end{thebibliography}
\end{document}